\documentclass[fleqn,preprint,5p]{elsarticle} 
\journal{Physics Letters B} 
\usepackage{xcolor} 
\definecolor{mygreen}{rgb}   {0.1,1.0,0.1}  
\definecolor{olive}{rgb}{1,1,0}                    
\definecolor{limegreen}{rgb} {0.243902,1.,0.243902}   
\usepackage{multicol}
\usepackage{times} 
\usepackage[hyphens]{url} 
\usepackage{hyperref}

\usepackage[anythingbreaks]{breakurl} 
\usepackage[utf8]{inputenc}
\usepackage{graphics}
\usepackage{epsfig}
\usepackage{amsmath,amsfonts,amssymb,amsthm}
\usepackage{booktabs}
\usepackage{slashed}

\newcommand{\DD}{\frac{d}{2}}

\newcommand{\bq}{ \begin {equation} }
\newcommand{\eq}{\end{equation}}
\newcommand{\ba}{\begin{eqnarray}}
\newcommand{\ea}{\end{eqnarray}}

\newcommand{\be}{\begin{equation}}
\newcommand{\ee}{\end{equation}}

\newcommand{\bea}{\begin{eqnarray}}
\newcommand{\eea}{\end{eqnarray}}

\def\mathswitch#1{\relax\ifmmode#1\else$#1$\fi}
\def\mathswitchr#1{\relax\ifmmode{\mathrm{#1}}\else$\mathrm{#1}$\fi}

\begin{document}
\begin{frontmatter}
\title{
\begin{flushright}
\tt{\small 
DESY 17-079
\\ 
HCMUS-19-01
\\[-2mm]
KW 18-004
}
\end{flushright}
\vspace*{0cm}
Scalar 1-loop Feynman integrals as meromorphic functions in space-time dimension~$d$
}

\author[label1]{Khiem Hong Phan}
\author[label2,label3]{Tord Riemann\corref{ca1}}
\cortext[ca1]{Corresponding author} 
\ead{tordriemann@gmail.com}

\address[label1]{University of Science, Vietnam National University, Ho Chi Minh City, Vietnam}
\address[label2]{DESY Deutsches Elektronen-Synchrotron, 15738 Zeuthen, Germany}
\address[label3]{Institute of Physics, University of Silesia, 40-007 Katowice, Poland}
\begin{abstract}
The long-standing problem of representing the general massive one-loop Feynman integral as a meromorphic function of the space-time dimension $d$ has been solved for the basis of scalar one- to four-point functions with indices one.
In 2003 the solution of difference equations in the space-time dimension allowed to determine the necessary classes of special functions: self-energies need ordinary logarithms and Gauss hypergeometric functions $_2F_1$,
vertices need additionally Kamp\'{e} de F\'{e}riet-Appell functions $F_1$,
and box integrals also Lauricella-Saran functions $F_S$.
In this study, alternative recursive Mellin-Barnes representations are used for the representation of $n$-point functions in terms of $(n-1)$-point functions.
The approach enabled the first derivation of explicit solutions for the Feynman integrals at arbitrary kinematics.
In this article, we scetch 
our new representations for the general massive vertex and box Feynman integrals and derive a numerical approach for the necessary Appell functions $F_1$ and Saran functions $F_S$ at arbitrary kinematical arguments.   
\end{abstract}
\end{frontmatter}

\allowdisplaybreaks
\section{Introduction    \label{s-i}}
We are studying scalar one-loop Feynman integrals,
\begin{align}
\label{npoint}
J_n (d)
 &=
    \int \dfrac{d^d k}{i \pi^{d/2}} \dfrac{1}{D_1^{\nu_1} D_2^{\nu_2}\cdots D_n^{\nu_n}}
,
 \end{align}
with inverse propagators $D_i= (k+q_i)^2-m_i^2+i\varepsilon$.
We assume $\nu_i=1$ as well as momentum conservation and all external momenta to be incoming, 
$\sum_{e=1}^np_e=0.$ 
The $q_i$ are loop momenta shifts and will be expressed for applications by the external momenta $p_e$.
Dimensions $d=4+2n-2\epsilon$ with $n\geq 0$ are of physical interest because tensor one-loop Feynman integrals of rank $r$ in $4-2\epsilon$ dimensions may be expressed by scalar integrals taken in higher dimensions up to $d=4+2r-2\epsilon$ 
\cite{Davydychev:1991va}.
Higher indices $\nu_i$ will also appear in the representation, but may be eliminated, so that a complete reduction basis  
of higher-dimensional scalar one- to four point integrals may be derived.  
For two- to seven-point tensor functions this has been worked out in \cite{Fleischer:2010sq,Riemann:september2013}.

The first terms of the $\epsilon$-expansion of one- to four-point scalar functions for $d=4-2\epsilon$, until including the constant term, was given by G. t'Hooft and M. Veltman in 1978
\cite{'tHooft:1978xw}.
A systematic numerical treatment of the next terms of order $\epsilon$-terms  
was performed in 1992 
\cite{Nierste:1992wg}, and a systematic numerical approach was worked out in 2001 
\cite{Passarino:2001wv}.
It has been shown in 2003 \cite{Tarasov:2000sf,Fleischer:2003rm} that
representations in general dimension $d$, including $d=4-2\epsilon$,  
will rely on certain multiple hypergeometric functions of the type
$_2F_1, F_1, F_S$.
Though, the explicit solutions for arbitrary kinematics could not be found. 

A scetch of the Feynman integrals at arbitrary kinematics in terms of $_2F_1, F_1, F_S$ and their explicit numerical determination are the subject of this letter. 
The dependence on the external momenta $p_e$ will be contained 
exclusively in the functions $R_n$:
\begin{eqnarray}\label{Rn}
R_n \equiv  R_{12\dots n}
=  -~\frac{\lambda_n}{G_n}  - i \varepsilon.
\end{eqnarray}
The $R_n$ carry the causal regulator $- i \varepsilon$.
The Cayley matrix $\lambda_{12\dots n}$ was introduced in \cite{Melrose:1965kb}.
It
is composed of the variables $Y_{ij}$,
and its determinant $\lambda_n$ is:
\begin{eqnarray} \label{npoint-cayley}
\lambda_n \equiv \det (\lambda_{12 \dots n})
&=&  
\left|
\begin{array}{cccc}
Y_{11}  & Y_{12}  &\ldots & Y_{1n} \\
Y_{12}  & Y_{22}  &\ldots & Y_{2n} \\
\vdots  & \vdots  &\ddots & \vdots \\
Y_{1n}  & Y_{2n}  &\ldots & Y_{nn}
\end{array}
         \right|
,
\end{eqnarray}
with
\begin{eqnarray}\label{eq-yij}
Y_{ij} = Y_{ji} &=& m_i^2+m_j^2-(q_i-q_j)^2 
. 
\end{eqnarray}
Further, we use the $(n-1)\times(n-1)$ dimensional Gram determinant $G_n$,
\begin{align}
 \label{Gram}
&
G_n \equiv -2^{n}~ \det(G_{12 \cdots n})
,
\end{align}
and
\begin{align}
 \label{Gram0}
& \det(G_{12 \cdots n}) =
 \\[1mm] \nonumber
& 
\hspace*{-5mm}
\left|
\begin{array}{cccc}
  \! (q_1-q_n)^2 
&\ldots & (q_1-q_n)(q_{n-1}-q_n) 
\\
  \! (q_1-q_n)(q_2-q_n) 
&\ldots 
& (q_2-q_n)(q_{n-1}-q_n) 
\\
  \vdots    
&\ddots   & \vdots 
\\
  \! (q_1-q_n)(q_{n-1}-q_n)    
&\ldots & (q_{n-1}-q_n)^2
\end{array}
\right|
.
\end{align}
We use the special assignment for tadpoles:
\begin{equation}
 G_1=-2
 .
\end{equation}
Both determinants $\lambda_n$ and $G_n$ are independent of a common shift of the internal momenta $q_i$.
Further, we introduce the notion $R(i)$,
\begin{equation}
 R(i) \equiv r(i) -i\varepsilon ~\equiv ~ -\det(\lambda_i)/G_1-i\varepsilon ~=~ m_i^2 -i\varepsilon
,
\end{equation}
and use, wherever it is unique from the context, 
\begin{equation}
R_1 \equiv R(i)
. 
\end{equation}

We derived in \cite{Bluemlein:2017rbi} a new ansatz, a recursion relation for the Feynman integrals defined in \eqref{npoint},
\begin{align} 
\label{JNJN1}
&
J_n (d) 
= \dfrac{{-1}}{2\pi i} \int\limits_{c_0-i\infty}^{c_0+i\infty}ds  
     \dfrac{\Gamma(-s) \Gamma(\frac{d-n+1}{2}+s)  }
                { 2\Gamma(\frac{d-n+1}{2}) } 
\nonumber 
\\
& \times       \Gamma(s+1)
   R_{n}^{-s-1}
\sum\limits_{k=1}^n 
             {\partial_k {R_n} } 
      \;
     {\bf k}^- J_n(d+2s) 
     ,
\end{align} 
and its solution by a sequence of Mellin-Barnes representations.
We use the representation $\partial_k {R_n}$ for the co-factor of the Cayley matrix, also called signed minors in 
e.g. \cite{Melrose:1965kb}:
\begin{align}\label{eq-partial}
 \partial_k {R_n} ~=~ \frac{\partial {R_n}}{\partial m_k^2} 
{ ~=~ 
  \left( 
\begin{array}{c}
0\\
k\\
\end{array}
\right)_n
}
.
\end{align}
The operator ${\bf k^{-}}$ reduces an $n$-point Feynman integral $J_n(d)$ to  $(n-1)$-point integrals $J_{n-1}(d+2s)$ 
by shrinking the $k^{th}$ propagator, $1/D_k$:
\begin{eqnarray}\label{eq-def-operator-k-}
{\bf k^{-}}~ J_n (d)
&=&
\int \dfrac{d^d k}{i \pi^{d/2}} \dfrac{1}{\prod_{j\neq k,j=1}^n D_j}
.
\end{eqnarray}
The recurrence relation \eqref{JNJN1} is the master integral for one-loop $n$-point functions in space-time dimension $d$, representing 
them by $n$ 
integrals over $(n-1)$-point functions with a shifted, continuous dimension $d+2s$.
The recurrence starts at $n=2$ with the tadpole $J_1(d)$ in the integrand:
\begin{align} \label{eq-tadpole}
&J_{1}(d;m_i^2) 
= 
\int \dfrac{d^d k}{i \pi^{d/2}} 
\dfrac{1}{k^2-m_i^2+i \varepsilon}
\nonumber\\
&=
- \frac{\Gamma( 1 -d/2)}
 {(m_i^2-i \varepsilon)^{1-d/2}}
 \equiv
 - \frac{\Gamma( 1 -d/2)}
 {{R_1}^{1-d/2}}
.
\end{align}
Eqn. \eqref{JNJN1} contains for $n=2$ the term {$\int ds (\frac{R_1}{R_2})^{s}$}, multiplied by 
$\Gamma$-matrices with arguments depending on $s$, and is
formally a \emph{Mellin-Barnes integral}.
Our representation is an alternative to Eq.~(19) of ~\cite{Fleischer:2003rm}. 
There, an {\it infinite sum} over a \emph{discrete dimensional parameter $s$} was derived in order to represent an $n$-point function $J_n(d)$ by integrals $J_{n-1}(d+2s)$. 

The further evaluations will depend, concerning the kinematics, exclusively on the $R_1, R_2,$ etc.
introduced in \eqref{Rn}.
Although, there will arise exceptional cases when the specific choi\-ce of the external scalars $(p_{e_i}p_{e_j})$ or of internal mass squares $m_i^2$ will lead to vanishing or divergent
determinants $\lambda_n$ or $G_n$.
In such cases, one has to go back to intermediate definitions and look for specific solutions.\footnote{A complete analysis of the exceptional kinematical cases has been performed by K.H.P; to be published elsewhere.}
See also the remarks in \cite{Usovitsch:2018shx}.

\section{Massive vertex and box functions \label{s-mvb}}
Representations of the massive self-energy, vertex and box integrals can be derived iteratively from \eqref{JNJN1} by closing the integration contours of the 
Mellin-Barnes integrals e.g. to the right and taking the two series of residues of the corresponding $\Gamma$-functions with arguments $(-s+\cdots)$.
One Cauchy sum constitutes the analogue of the so-called \emph{boundary or $b$-terms} of \cite{Fleischer:2003rm}, the other one has a genuine $d$-dependence.
Both sums together represent the Feynman integrals.
In our approach, closed analytical expressions could be determined for arbitrary kinematics.

The general massive vertex and box integrals $J_3(d), J_4(d)$ have first been published at LL2018 \cite{Riemann:April2018}.
An alternative, instructive version of the vertex is
 \begin{eqnarray}\label{J3}
J_3(d)= {J_{123}} + J_{231} + J_{312}
,
\end{eqnarray}
with short notations $R_3=R_{123}, R_2=R_{12}$  etc., and:
\begin{align}  
\label{J123altern}
&J_{123}  
=
\Gamma\left( 2-\frac{d}{2} \right)  
{ \frac{\partial_3 R_3}{R_3} \frac{\partial_2 R_2}{R_2}  }
\frac{ {R_2}} 
{2\sqrt{1-{R_1}/R_2}}
\\[3mm] \nonumber                   
&
\Biggl[ 
- ~
R_2^{\frac{d}{2}-2}  
\frac{\sqrt{\pi}}{2} 
\frac{\Gamma\left(\frac{d}{2}-1\right) } {\Gamma\left(\frac{d}{2}-\frac{1}{2}\right) }
~ _2F_1 \left( \frac{d-2}{2},1; \frac{d-1}{2}; \frac{R_{2}}{R_{3}} \right)
\\ \nonumber
& +~ R_3^{\frac{d}{2}-2}  
~_2F_1 \left(1,1;\frac{3}{2}; \frac{R_{2}} {R_{3}} \right)
\Biggr]   
\nonumber
\\[3mm]     \nonumber
&               
+ ~
\Gamma \left(2-\frac{d}{2} \right)  
{
\frac{\partial_3 R_3} {R_3}   
\frac{\partial_2 R_2}{R_2} 
}
 \frac{{R_1} } {4\sqrt{1-{R_1}/R_2}} 
 \nonumber
\\[3mm]                    
&
 \Biggl[
+
\frac{2 {R_1}^{\frac{d}{2}-2}  }{d-2}
F_1 \left( \dfrac{d-2}{2}; 1, \frac{1}{2}; \DD; \dfrac{{R_1}}{R_{3}}, \dfrac{{R_1}}{R_{2}} \right)
\nonumber \\ \nonumber
&
-~ 
R_3^{\frac{d}{2}-2}  
F_1 \left( 1; 1, \frac{1}{2}; 2; \dfrac{{R_1}}{R_{3}}, \dfrac{{R_1}}{R_{2}} \right)
\Biggr]  
\nonumber
\\[3mm]                     
&
+ ~~ ({R_1(1)} \leftrightarrow {R_1(2)})  
.
  \nonumber
\end{align}
We use the abbreviation \eqref{eq-partial}.
For $d \to 4$, 
both the sums of expressions with $_2F_1$ and $F_1$ 
in square brackets in \eqref{J123altern} approach zero, thus compensating the pole factor
$\Gamma(2-{d}/{2})$ in this limit.
The $J_3$ stays finite at $d=4$, as it should be for any massive 3-point function. 
And the $\epsilon$ expansion for $J_{123}$ to order $n$ needs, in this case, the evaluation of the components to order $(n+1)$.

The {corresponding} massive four-point function is:
 \begin{eqnarray}\label{J4}
 J_4(d)= {J_{1234}} + J_{2341} + J_{3412} + J_{4123}
 ,
\end{eqnarray}
with $R_4=R_{1234}, R_3=R_{123}, 
R_2=R_{12}$ etc., and:
\begin{align} \label{J1234altern}%
&
J_{1234} 
=  
  \Gamma\left(2-\frac{d}{2}\right)  
 { \frac{\partial_4 R_4}{R_4}}
    \Biggl\{
\nonumber
\\                
&
\Biggl[ 
\frac{b_{123}}{2}  
\Biggl(
- ~
R_3^{\frac{d}{2}-2}  
~_2F_1 \left( \frac{d-3}{2},1; \frac{d-2}{2} ; \frac{R_{2}}{R_{3}} \right)
\nonumber \\ \nonumber
&+~  R_4^{\frac{d}{2}-2}  
~ \sqrt{\pi} ~ 
\frac{\Gamma\left(\frac{d-2}{2}\right) }{\Gamma\left(\frac{d-3}{2}\right) }
\nonumber \\ \nonumber
& \times
~_2F_1 \left( \frac{1}{2},1; 1 ; \frac{R_{2}}{R_{3}} \right)
\Biggr)\Biggr]   
\nonumber
\\   \nonumber                 
&
+ ~
\frac{\Gamma\left(\frac{d-2}{2}\right) }{\Gamma\left(\frac{d-3}{2}\right) }
~ \frac{\sqrt{\pi}}{4} ~
{ 
\frac{\partial_3 R_3}{R_3} 
}
\frac{
{ 
\partial_2 R_2}
}{\sqrt{1-{R_1}/R_2}} 
\nonumber \\ \nonumber
&  \times
~ {_2F_1 \left( \frac{1}{2}, 1; 1 ; \frac{R_{2}}{R_{3}} \right)}
\\                     
&
 \Bigl[
+ ~
\frac{R_2^{\frac{d}{2}-2} }{d-3}
F_1 \left( \dfrac{d-3}{2}; 1, \frac{1}{2}; \dfrac{d-1}{2} ; \dfrac{R_2}{R_{4}}, \dfrac{R_2}{R_{3}}\right)
\nonumber \\ \nonumber
&-~   
R_4^{\frac{d}{2}-2}  
 F_1 \left( \dfrac{1}{2}  ; 1, \frac{1}{2}; \dfrac{3}{2}   ; \dfrac{R_2}{R_{4}},\dfrac{R_2}{R_{3}}\right)
\Bigr]  
  \nonumber \\    \nonumber               
&
+ ~
\frac{{R_1}}{8} ~
\frac{\Gamma\left(\frac{d-2}{2}\right) }{\Gamma\left(\frac{d-3}{2}\right) }
{
\frac{\partial_3 R_3}{R_3}   
\frac{\partial_2 R_2}{R_2} 
\frac{1}{1-{R_1/R_3}}
\frac{1}{1-{R_1/R_2}}
}
\nonumber \\                    
\hspace*{-2cm}
&
\Bigl[
- ~
{R_1^{\frac{d}{2}-2}} 
\frac{\Gamma\left(\frac{d-3}{2}\right) }{\Gamma\left(\frac{d}{2}\right) }
\nonumber \\   \nonumber
&
\times  \hspace*{-0.5mm} 
F_S \Biggl
( \frac{d-3}{2},1,1;1,1,\frac{1}{2};\frac{d}{2},\frac{d}{2},\frac{d}{2};\frac{{R_1}}{R_4},
\cdots
\nonumber \\   \nonumber
&
\hphantom{F_S \Biggl( d/2-3/2,1,1;1,1,}
\frac{{R_1}}{{R_1}-R_3},\frac{{R_1}}{{R_1}-R_2} \Biggr)
\nonumber \\   \nonumber               
&
+ ~
R_4^{\frac{d}{2}-2} ~ \sqrt{\pi} 
\nonumber \\   \nonumber
&
\times \hspace*{-0.5mm} 
F_S(\frac{1}{2},1,1; 1,1, \frac{1}{2} ;2,2,2,\frac{{R_1}}{R_4},\frac{{R_1}}{{R_1}-R_3},\frac{{R_1}}{{R_1}-R_2})
 \Bigr] 
\\\nonumber 
& + ~ ({R_1(1)}   \leftrightarrow {R_1(2)} ) 
\Biggr\}
\\ 
& +~ (2,3,1) + (3,1,2) 
,
\end{align} 
where the function $b_{123}$ is independent of $d$,
\begin{align}
\label{b123normal}
\nonumber 
&b_{123} 
=   \frac{1}{2}
\frac{ \partial_3 R_{3} }{R_{3}} 
\frac{ \partial_2 R_{2} }{R_{2}} 
\Biggl[
\frac{R_2}{ \sqrt{ 1- \frac{ {R_1}}{R_{2}}  }  }  
~  _2F_1 \left(1, 1; \frac{3}{2} ; \frac{R_{2}}{R_{3}} \right)
\\
&
-~ 
\frac{1}{2}
\frac {R_1} { \sqrt{1-\frac{{R_1}}{R_{2}} }} 
 ~ F_1 \left( 1; 1, \frac{1}{2}; 2; 
              \frac{ {R_1} }{R_{3}}, \frac{ {R_1} }{R_{2}} \right)
\Biggr]
 +  (1 \leftrightarrow 2)      
.
\end{align}
Here, it is $R_1=R_1(1)$ and \eqref{eq-partial} defines derivatives like $\partial_2 r_2$.
The term $b_{123}$, when 
multiplied with $\Gamma(-\frac{d-4}{2}) R_3^{\frac{d}{2}-2}$,
equals the term of $J_{123}$ in \eqref{J123altern} with $d$-in\-de\-pen\-dent $F_1$ and $F_S$.
It replaces the so-called $b_3$-term of the vertex integral in \cite{Fleischer:2003rm} for arbitrary kinematics, while
the $d$-dimen\-sio\-nal parts of $J_{1234}$ agree.

For $d \to 4$, all the expressions in square brackets in \eqref{J1234altern} approach zero, thus compensating the pole of $\Gamma(2-{d}/{2})$ in this limit. 
As a result, the $J_4$ stays finite at $d=4$, as it should be for any massive 4-point function. And the $\epsilon$ expansion for $J_{1234}$ to order $n$ needs, in this case, the evalution of the components to order $(n+1)$.

 The derivations of $J_{123}$ and $J_{1234}$ were done under the assumption that the kinematical arguments $x,y,z$ of the $_2F_1, F_1$, $F_S$ fulfill $|x|,|y|,|z| <1 $.
Nevertheless, the above formulae  are valid  at arbitrary kinematical arguments, 
for massive vertices at $\Re e(d)>2$ and for box integrals  at $\Re e(d)>3$.
In ~\ref{s-app} to~\ref{app-subs-fs} 
we will show how to calculate the various $F_1$ and $F_S$ for arbitrary complex arguments; for $_2F_1$ we assume that such calculations are well-known.

\section{%
Numerical results \label{sec-numerics}}
The scalar one-loop basis consists of one- to four-point functions.
Our two-point function $J_2(d)$ was reproduced in \cite{Riemann:April2018} and is in complete agreement with \cite{Fleischer:2003rm}, while for $J_3(d)$ and $J_4(d)$ our results are novel.
Concerning numerical results for the 3-point functions we refer to 
several tables in \cite{Riemann:2017podlesice,Bluemlein:2017rbi}.
The kinematics was chosen such that the results of \cite{Fleischer:2003rm} could be compared.\footnote{We would like to thank Oleg Tarasov for a helpful discussion concerning this issue.}
Another numerical comparison, for a box integral $J_4(d)$ with vanishing Gram determinant, may be found in \cite{Usovitsch:2018shx,Usovitsch:January2018}.

In Table \ref{tab-D0-1}  we show few examples of four-point functions in comparison to other packages.
We did not aim at maximal accuracy and claim essentially eight safe digits.  
Further, one propagator is massive and $d=4$ or $d=5$, and we can also allow for complex masses at the internal lines.
A true sample $\varepsilon$-expansion is reproduced for the generalized hypergeometric function $F_1$ in Table \ref{t-f1numereps}.

\begin{table}[t]                
\caption{\label{tab-D0-1}
Comparison of 
the box integral $J_4$ defined in \eqref{J1234altern} with the LoopTools
function 
{\tt D0($p_1^2, p_2^2, p^2_3, p^2_4, (p_1+p_2)^2, (p_2+p_3)^2, m^2_1, m^2_2, m^2_3 , m^2_4$)}
\cite{Hahn:1998yk,vanOldenborgh:1990yc} 
at 
$m^2_2= m^2_3= m^2_4=0$.
Further numerical references are the
packages K.H.P\_D0 (PHK, unpublished) 
and MBOneLoop \cite{Usovitsch:January2018}.
External invariants:
$(p_1^2=\pm 1,p_2^2=\pm 5,p^2_3=\pm 2,p^2_4=\pm 7,s=\pm 20,t=\pm 1)$.
}
\begin{center}
\renewcommand{\arraystretch}{1.2}
\begin{tabular}{|l|l|} \hline \hline
$(p_1^2,p_2^2,p_3^2,p_4^2, s, t)$ 
                           & 4-point integral   
\\ \hline 
$(-,-,-,-,-,-)$ & $d=4$, ~ $m^2_1=100$
\\
$J_4$ & $0.00917867$ 
\\
LoopTools                     &
 $0.0091786707$ 
\\ MBOneLoop &  $0.0091786707$
\\ \hline
$(+,+,+,+,+,+)$       & $d=4$, ~ $m^2_1=100$
\\
$J_4$ & $-0.0115927 - 0.00040603\;i$   
\\
LoopTools                      & $-0.0115917- 0.00040602\;i$ 
\\ MBOneLoop & $-0.0115917369 - 0.0004060243 
\;i$
\\ \hline \hline
$(-,-,-,-,-,-)$ & $d=5$, ~ $m^2_1=100$
\\ 
$J_4$ & $ 0.00926895  $ 
\\
K.H.P\_D0                    &
 $0.00926888 $
\\ MBOneLoop & $0.0092689488 
$
\\ \hline
$(+,+,+,+,+,+)$       & $d=5$, ~ $m^2_1=100$
\\
$J_4$ & $-0.00272889 + 0.0126488\;i$
\\
K.H.P\_D0                   & \hspace*{+19.5mm}(--)
\\ MBOneLoop & $-0.0027284242 +  0.0126488134 
\;i$
\\ \hline \hline
$(-,-,-,-,-,-)$ & $d=5$, ~ $m^2_1=100-10~i$
\\
$J_4$ & $ 0.00920065 + 0.000782308~i$
\\
K.H.P\_D0                    &
 $0.0092006 ~~ +0.000782301~i$
\\ MBOneLoop & $ 0.0092006481 + 0.0007823090\;i$
 \\ \hline
$(+,+,+,+,+,+)$       & $d=5$, ~ $m^2_1=100-10~i$
\\
$J_4$ & $-0.00398725 + 0.012067\;i$
\\
K.H.P\_D0                   & $-0.00398723+ 0.012069\;i$
\\ MBOneLoop & $-0.0039867702 + 0.0120670388 
\;i$
\\ \hline\hline
\end{tabular}
\end{center}
\end{table}

For the safe numerical calculation of massive vertices $J_3$ and massive box integrals $J_4$ we collect stable numerical representations for the generalized hypergeometric functions $F_1$ and $F_S$ in the Appendices.

\section{%
Discussion \label{sec-discussion}}
The massive oneloop Feynman integrals have been represented as meromorphic functions of space-time $d$ in terms of generalized hypergeometric functions.
Many details left out here will be published elsewhere.
The Feynman integrals can be calculated numerically at arbitrary kinematics and arbitrary dimension $d$, including potential pole locations at $d=4+2n$. 
For phenomenological or multi-loop applications, it is wishful to have the pole expansions in closed analytical form.
Their derivation is subject of a subsequent study.

The new recursion \eqref{JNJN1} has a unique feature.
It allows to derive $n$-dimensional Mellin-Barnes integrals for $n$-point Feynman integrals.
Generally, $n$-dimensional integrals are obtained by sector decomposition methods, while in the Mellin-Barnes approach, as it is advocated in numerical loop calculations, the number of dimension grows faster.
Within the MBsuite, AMBRE generates for the most general massive $n$-point one-loop function an $N_n=\frac{1}{2}n(n+1)$-dimensional MB-integral; according to the number of entries $Y_{ij}$ in the second Symanzik polynomial,
$F(x)=\frac{1}{2} x_iY_{ij}x_j-i\varepsilon$. For a vertex or box, $N_{3}=6, N_4=10$.
In the present approach, it is only $N_{3}^{\prime}=3, N_{4}^{\prime}=4$.
Evidently, a replacement of the original kinematical invariants $m_i^2, (p_{e,i}p_{e,j})$ or $Y_{ij}$ by the alternatives
$R_n=-\lambda_n/G_n$ is an essential building block and it might well be possible to find similar lower-dimensional MB-representations also for more involved multi-loop integrals.

Basic numerical features of the new $n$-dimensional MB-repre\-sen\-ta\-tion \eqref{JNJN1} have been studied in 
\cite{Usovitsch:2018LLslidesJU,Usovitsch:January2018} in comparison with \cite{Fleischer:2010sq}, 
with the package 
MBOne\-Loop,including cases of small or vanishing Gram determinant. 

It is interesting to compare our results for  $J_3(d)$ and $J_4(d)$ with the earlier study \cite{Fleischer:2003rm}.
The $d$-dependent part of $J_3(d)$ as well as much of the $d$-dependent part of 
$J_4(d)$ agree with our results. 
Further, the expressions for the $b$-terms in \cite{Fleischer:2003rm} differ from our 
$d$-independent parts, although in certain kinematical regions they do agree numerically for $J_3(d)$.
We find no agreement for $J_4(d)$, due to the various contributing $b$-terms.

\appendix
\setcounter{table}{0}

\section{The Appell function~$F_1$ and Lauricella-Saran function~$F_S$
\label{s-app}
}
Numerical calculations of specific Gauss hypergeometric functions $_2F_1$, 
Appell functions $F_1$ (Eqn. (1) of \cite{Appell:1925}), and Lauricella-Saran functions $F_S$ (Eqn. (2.9) of \cite{Saran:1955}) are needed for the scalar one-loop Feynman integrals:
\begin{align}
 \label{eq-2f1def}
& _2F_1(a,b;c;x)
 =
\sum_{k=0}^{\infty} \frac{(a)_k(b)_k}{ k! ~ (c)_k} ~ x^k,
\\\label{eq-f1def}
& F_1(a;b,b';c;y,z) 
\nonumber\\
&=
\sum_{m,n=0}^{\infty} \frac{(a)_{m+n} (b)_m (b')_n}{ m! ~ n! ~ (c)_{m+n}} ~ y^m z^n,
\\\label{eq-fsdef}
& F_S(a_1,a_2,a_2; b_1,b_2,b_3 ; c,c,c ; x,y,z)
\\\nonumber
&=
\sum_{m,n,p=0}^{\infty} \frac{(a_1)_{m} (a_2)_{n+p} (b_1)_m (b_2)_n (b_3)_p}
{ m! ~ n! ~ p! ~(c)_{m+n+p}} ~ x^m y^n z^p
.
\end{align}
The  $(a)_k$ is the Pochhammer symbol.
The series converge for $|x|, |y|, |z| <1$, but the functions are needed for arbitrary arguments.
All the $_2F_1, F_1, F_S$ are finite and have no pole terms in $\epsilon$.
Practically all aspects of  $_2F_1$ are well-known and implemented in computer algebra systems, in Mathematica as built-in symbol {\tt Hypergeometric2F1[a,b,c,z]}.
There is no public $F_S$-package, while the Appell function 
$F_1(a;b_1,b_2;\linebreak[4]c;x,y)$ \cite{Appell:1925} is implemented in  Mathematica
as built-in symbol {\tt AppellF1[a,b1,b2,c,x,y]}
\cite{appell:mathworld}.
Another public package is {\tt f1} \cite{Colavecchia:2001,Colavecchia:x2004}, and a wrapper package for  {\tt f1} is {\tt appell} \cite{DanielSabanesBove:2015}.
All the implementations mentioned have systematic limitations.

One approach to the numerics of $F_1$ and $F_S$ may be based on Mellin-Barnes representations.
For the Gauss function $_2F_1$ and the Appell function $F_1$, Mellin-Barnes representations are known.
See Eqn.~(1.6.1.6) in \cite{Slater:1966},
\begin{align}\label{s-mb2f1}
& _2F_1(a,b;c;z) = 
\frac{1}{2\pi i}~
\frac{\Gamma(c)}{\Gamma(a)\Gamma(b)}
\\\nonumber
& \times \int_{-i\infty}^{+i\infty} 
ds ~ (-z)^s ~ \frac{\Gamma(a+s)\Gamma(b+s)\Gamma(-s)}{\Gamma(c+s)}
,
\end{align}
and Eqn.~(10) in \cite{Appell:1925}, which is a two-dimensional MB-integral:
\begin{align}\label{s-mbf1}
&F_1( a;b,b';c;x,y) = 
\frac{1}{2\pi i}~
\frac{\Gamma(c)}{\Gamma(a)\Gamma(b')}
\\\nonumber
&
\times
\int_{-i\infty}^{+i\infty} dt  ~ (-y)^t ~ _2F_1(a+t,b;c+t,x)
\\\nonumber
&
\times \frac{\Gamma(a+t)\Gamma(b'+t)\Gamma(-t)}{\Gamma(c+t)}
.
\end{align}
For the Lauricella-Saran function $F_S$ we derived the following, new, three-dimen\-sio\-nal MB-integral: 
\begin{align}\label{s-mbfs}
&F_S(a_1,a_2,a_2;b_1,b_2,b_3;c,c,c;x,y,z) 
\\\nonumber
&
= 
\frac{1}{2\pi i}~
\frac{\Gamma(c)}{\Gamma(a_1)\Gamma(b_1)}
\int_{-i\infty}^{+i\infty} dt 
~ F_1(a_2;b_2,b_3;c+t;y,z)
\\\nonumber
&
\times  (-x)^t ~ \frac{\Gamma(a_1+t)\Gamma(b_1+t)\Gamma(-t)}{\Gamma(c+t)}
.
\end{align}
A general numerical evaluation of these representations deserves some sophistication.
Let us mention the simple one-loop massive QED vertex for which no trivial MB method exists when the kinematics is Minkowskian, a problem discussed e.g. in  \cite{Czakon:2005rk} and solved in \cite{Dubovyk:2016ocz}. 
It was demonstrated in \cite{UsovitschRiemannMinirep:2018aa} that MB\-One\-Loop, a fork of the package MBnumerics 
\cite{Usovitsch:201805-phan-riemann,Usovitsch:2018shx,Usovitsch:January2018} may be used to solve \eqref{s-mb2f1} to \eqref{s-mbfs} at arbitrary kinematics with high precision. 

One might also try to approach the generalized hypergeometric functions using Pochhammer's double loop contours \cite{Pochhammer:1870,Erdelyi:1950}, or study the defining differential equations \cite{Bytev:2011ks,Bytev:2013bqa,Bytev:2013gva}, etc.
After several trials, we decided to base our numerics on the integral representations of $F_1$ proposed in \cite{Picard:1881} and $F_S$ proposed in \cite{saran1955}; see \ref{app-subs-f1} and \ref{app-subs-fs}.
Astonishing enough, it will (nearly) suffice to use mathematics known to well-educated German gymnasiasts.

\section{The Appel functions $F_1$ 
\label{app-subs-f1}}
A one-dimensional integral representation for $F_1$ ~\cite{Picard:1881}
is quoted in Eqn.~(9) of \cite{Appell:1925}:
\begin{align}\label{s-f1A}
& F_1(a;b,b';c;x,y)
 =
 \frac{\Gamma(c)}{\Gamma(a)\Gamma(c-a)}
 \\\nonumber
 & \times
 \int_{0}^{1}du\frac{u^{a-1}(1-u)^{c-a-1}}{(1-xu)^b(1-yu)^{b'}}
 .
\end{align}
 We need three specific cases, taken at $d \geq 4$.
Namely for vertices:
\begin{align}\label{F1-002}
& F_1^v(d)
\equiv F_1 \left( \frac{d-2}{2};1,\frac{1}{2};\frac{d}{2};x_c,y_c \right)
\\\nonumber
&=    
\frac{1}{2}(d-2) \int_0^1 \frac{ du~u^{\frac{d}{2}-2}}{(1-x_c u)\sqrt{1-y_c u}}
.
\end{align}
Integrability is violated at $u=0$ if not $\Re e(d)>2$.
Similarly, for box integrals:
\begin{align}\label{F1-003}
& F_1^b(d)
\equiv
F_1 \big(\frac{d-3}{2};1,\frac{1}{2};\frac{d-1}{2};x_c,y_c\big)
\\\nonumber
&=
\frac{1}{2}(d-3)\int_0^1 \frac{ du~u^{d/2-5/2}}{(1-x_c u)\sqrt{1-y_c u}}
\\\nonumber
&= F_1^v (d-1) 
,
\end{align}
Integrability is violated at $u=0$ if not $\Re e(d)>3$.
Finally for the definition of the box Saran function $F_S$ \eqref{fsgeneral}:
\begin{align}\label{F1sbox}
& F_1^S{(y_c,z_c)}
 \equiv
 F_1(1;1,\frac12,\frac32;y_c,z_c)
 \\\nonumber
&
= \frac12 \int_0^1 \frac{{\color{black} u}~du}{\sqrt{1-u} (1-x_cu)\sqrt{1-y_cu}}
.
 \end{align}
The singularity at $u=1$ is integrable.

\subsection{Specific values of $_2F_1$ and $F_1$  at $d=4$ 
\label{ss-dfour}}
The vertex function \eqref{J123altern} contains $_2F_1$ and $F_1$ with specific values at $d=4$:
\begin{eqnarray}\label{2F1limit3}
_2F_1\left(1,1;\frac{3}{2};x_c\right)
&=& 
\frac{ \mathrm{ArcSin}(\sqrt{x_c}) }{\sqrt{1 - x_c} \sqrt{x_c}}
\end{eqnarray}
and
\begin{align}
\label{F1limit3}
&F_1 \left( 1; 1, \frac{1}{2}; 2; x_c,y_c \right)
=
2 \frac{
\mathrm{ArcTanh}\left[\frac{\sqrt{x_c} \sqrt{1 -  y_c}}{\sqrt{x_c - y_c}}\right]} {\sqrt{x_c} \sqrt{x_c - y_c}}
\nonumber
\\
&
-
2 \frac{
\mathrm{ArcTanh}\left[\frac{\sqrt{x_c}}{\sqrt{x_c - y_c}}\right]} {\sqrt{x_c} \sqrt{x_c - y_c}}
.
\end{align}
Using logarithms only, $\mathrm{ArcSin}(z)=-i\ln(iz+\sqrt{1-z^2})$ and
$\mathrm{ArcTanh}(z) = \frac{1}{2}[\ln(1+z)-\ln(1-z)]$.
Eqn. \eqref{F1limit3} is only valid if $(x_c-y_c)$ has a well-defined imaginary part.
For $x_c = x-i\varepsilon_x$ and $y_c=y-i\varepsilon_y$ this is not necessarily the case
if $\varepsilon_x$ and $\varepsilon_y$ are independent and both infinitesimal.
So \eqref{F1limit3} has to be used with a grain of care.

The box function \eqref{J1234altern} contains additional $_2F_1$ and $F_1$ with specific values at $d=4$:
\begin{eqnarray}\label{2F1limita}
_2F_1 \left( \frac{1}{2}, 1; 1 ; x_c \right)
&=&                              
\frac{1} {\sqrt{1 - x_c} }
\end{eqnarray}
and 
\begin{align} \label{J1234aux2}
\hspace*{-7mm}
F_1 \left( \frac{1}{2}  ; 1, \frac{1}{2}; \frac{3}{2}   ; x_c,y_c  \right)
= \frac{1}{\sqrt{1-y_c}} ~~ _2F_1(\frac{1}{2};1,\frac{3}{2};\frac{x_c-y_c}{1-y_c})
\nonumber\\
 =
   \frac{\mathrm{ArcTanh}\left( \sqrt{\frac{x_c-y_c}{1-y_c}} \right) }  {\sqrt{x_c-y_c}}
.
\hspace*{14mm}
\end{align}
Eqn. \eqref{J1234aux2} is only valid if $(x_c-y_c)$ has a well-defined imaginary part.
Finally, we like to mention that we have no analogue to \eqref{2F1limita} and \eqref{J1234aux2}
for $F_S$ at $d=4$, namely
$F_S(\frac{1}{2},1,1; 1,1,\frac{1}{2}; 2,$ $2,2; x_c,y_c,z_c)$.

The 
Appell function $F_1^S = F_1(1;1,\frac12;\frac32;y_c,z_c)$ used in the integrand of the definition of the  Saran function \eqref{fsgeneral} can also be simplified:
\begin{align}\label{fs1dim}
\hspace*{-7mm}
F_1(1;1,\frac{1}{2};\frac{3}{2};y_c,z_c) 
=
\frac{1}{1-z_c}  ~
{_2F_1}\left( 1,1;\frac{3}{2}; \frac{y_c-z_c}{1-z_c}  \right) 
\nonumber\\
= \frac{\mathrm{ArcSin} \sqrt{\frac{y_c-z_c}{1-z_c}} }         {\sqrt{(y_c-z_c)(1-z_c)}}
.\hspace*{14mm}
\end{align}
Both representations in \eqref{fs1dim} are only valid when the imaginary part of the difference $(y_c-z_c)$ is well-defined.

For the Feynman integrals studied here, we have to take into account that $x_c, y_c$ and $z_c$ may have, in general, 
{\em uncorrelated infinitesimal} imaginary parts, and so their difference may be {\em not} well-defined.
Let us remind that  
$x_c = R_1/R_4$, and 
$y_c = R_{1}/(R_1 - R_3)$, and
$z_c = R_{1}/(R_1 - R_2)$.
Here, all the $R_n$ have, according to \eqref{Rn}, identical imaginary parts $- i \varepsilon$.
This leads to different infinitesimal imaginary parts 
$-\varepsilon_x,-\varepsilon_y,-\varepsilon_z$, with potentially different signs.
So, one has basically two equivalent options.
Either one treats $\varepsilon_x, \varepsilon_y$ and $\varepsilon_z$ as independent quantities and avoids the appearance of terms like $(x_c-y_c)$ and $(y_c-z_c)$.
Or one uses the exact knowledge of the imaginary parts of the $R_n$ from their definitions and arrives at well-defined imaginary parts of these   
$(x_c-y_c)$ and $(y_c-z_c$).

\subsection{Numerical calculation of $F_1^v(d)$ \label{ss-f1vxy1}}
For $x_c=x-iX$ and $y_c=y-iY$, Eqn.~ \eqref{F1-002} may be used for numerics if
$(X,Y) \geq \mathrm{const.}>0$ or if $(x,y)<1$.
The remaining cases $(X=-\varepsilon_x,Y=-i\varepsilon_y)\to +0$ 
deserve a closer inspection. They appear from Feynman integrals.

We exemplify here the first one of the two more involved cases: $1<x<y$ and $1<y<x$.

We introduc an auxiliary split parameter
\begin{equation} \label{eq-splitum}
 u_m= \frac{1}{2} \left( \frac{1}{y} + \frac{1}{x} \right)
\mathrm{~with~} 0 < \frac{1}{y} < u_m < \frac{1}{x} < 1
.
 \end{equation}
In the integrand of $F_1$ there will a cut begin at $u=\frac{1}{y}$ and  a pole arise 
at $u=\frac{1}{x}$ for infinitesimal $\varepsilon_x, \varepsilon_y$.
A split of the integral at $u_m$, 
\begin{equation}\label{eq-split}
 \int_0^1 du = \int_0^{u_m} du  + \int_{u_m}^1 du \equiv i_L + i_R 
 ,
\end{equation}
will lead to a separation of the singularities.
In both integrals at the right hand side, the integrand is regular with {\em one} exclusion.
We discuss now several opportunities of calculations, all of them with an accuracy of eight safe digits or better.

Our most careful approach persued the following ansatz with additional splittings:
 \begin{align}
 & F_1^v(d) 
 =
 I_A + I_0 + I_C + I_D + I_B + I_E
 \nonumber \\ &
 = 
 \lim_{R\to +0} 
 \Biggl[
 \int_{0}^{\frac{1}{y}-R} 
 + \int_{\frac{1}{y}-R}^{\frac{1}{y}+R} 
 + \int_{\frac{1}{{y}}+R}^{u_m} 
\nonumber \\ &
 + \int_{u_m}^{\frac{1}{x}-R} 
  + \int_{\frac{1}{x}-R}^{\frac{1}{x}+R} 
  + \int_{\frac{1}{x}+R}^{1} 
 \Biggr]
 \end{align}
After performing the limit $R\to0$ wherever possible, the integrals $A$ and $B$ will give real contributions, and the others are purely imaginary:
\begin{eqnarray} \label{eq-A0CDBE} 
   F_1^v(d) &=& 
      [ \Re e F_1^v(d)]  + i~[ \Im m F_1^v(d)] 
   \nonumber \\ 
   &=&
  \left[ A+ \mathrm{sign}(\varepsilon_x) \mathrm{sign} (\varepsilon_y)~B \right]
  \nonumber \\ &&
  +~ 
i~ \left[  \mathrm{sign} (\varepsilon_y)~(-C+D+E) \right]
.
\end{eqnarray}
It is 
\begin{align} 
\label{eq-A} 
& I_0=0 
,
\\ 
&
A=       \frac{d-2}{2} ~    \int_0^{\frac{1}{y}}
 \frac{ du ~ u^{d/2-2} } { (1-xu)\sqrt{1-yu}} 
, 
\\ &
B= \frac{d-2}{2} ~ \pi ~ \frac{1} { x \sqrt{\frac{y}{x}-1} ~ x^{d/2-2} }
,
\\ &
C= 
\frac{d-2}{2} ~
\int_{\frac{1}{y}}^{u_m}  \frac{du  ~ u^{d/2-2}} {(1-xu)\sqrt{yu-1}}
,
\\ & \label{eq-D}
D= 
\frac{d-2}{2}
\int_{u_m}^{\frac{1}{x}} \frac{du}{1-xu} 
\nonumber \\ &
\left( 
\frac{u^{d/2-2}}{\sqrt{yu-1}} - \frac{{x}^{-d/2+2}}{\sqrt{\frac{y}{x}-1}}
\right)
\nonumber \\ &
+
\frac{d-2}{2}
\frac{1}{\sqrt{\frac{y}{x}-1} ~ x^{d/2-2}} 
\nonumber \\ &
\left[ \ln(R) - \ln(\frac{1}{2x}-\frac{1}{2y}) \right] 
,
\\ &  \label{eq-E}
 E=
 \frac{d-2}{2}
\int_{\frac{1}{x}}^1 \frac{du}{1-xu} 
\nonumber \\ &
\left( 
\frac{u^{d/2-2}}{\sqrt{yu-1}} - \frac{{x}^{-d/2+2}}{\sqrt{\frac{y}{x}-1}}
\right)
\nonumber \\ &
+
\frac{d-2}{2}
\frac{1} {\sqrt{\frac{y}{x}-1} ~ x^{d/2-2}} 
\nonumber \\ &
\left[- \ln(R) + \ln(1-\frac{1}{x}) \right] 
.
\end{align}
 The remaining $R$-dependences in \eqref{eq-D} and \eqref{eq-E} drop out in the sum of $D$ and $E$.
 
 \medskip
 
Alternatively, with a subtraction in each of the two partial integrals
in \eqref{eq-split}, one may regularize the integrand of  $F_1^{v}(d)$ as follows:
\begin{eqnarray}
 i_L 
 &=& 
 \int_0^{u_m} du ~ \frac{g_x(u) - g_x\left(\frac{1}{y}\right)}  {\sqrt{1-y u}}
 + i_L^{ana}
 ,
 \end{eqnarray}
 \begin{eqnarray}
   i_R &=& \int_{u_m}^1 du ~ \frac{g_y(u) - g_y\left(\frac{1}{x}\right)}  {1-x u}
+ i_R^{ana}
,
\end{eqnarray}
 with
\begin{align}
&
i_L^{ana}
 =
-~2 ~ \frac{g_x\left(\frac{1}{y}\right)}{y_c}   \left[ \sqrt{1-y_c u_m} - 1  \right]
 \\\nonumber
 & \to
 - ~ 2 ~ \frac{g_x\left(\frac{1}{y}\right)}{y_c}   \left[ - 1 +i~ \mathrm{sign}(\varepsilon_y) \sqrt{y u_m-1} \right]
,
\end{align}
\begin{align}
&
i_R^{ana}
=
- \frac{g_y\left(\frac{1}{x}\right)}{x_c}
\ln\left(\frac{1-x_c}{1-x_c u_m}  \right)
 \\\nonumber
 & 
\to
- \frac{g_y\left(\frac{1}{x}\right)}{x}
\Biggl[
\ln\left( \frac{x-1}{1-x u_m}\right) 
{+}~ i\pi ~ \mathrm{sign}\left(\varepsilon_x\right)
\Biggr]
.
\end{align}

\medskip

Finally, a simplest approach will also do a reasonable numerics:
Perform  mean value integrals, like e.g. the built-in function of Mathematica:
\begin{align}
\hspace*{-5mm}
F_1^v(d) 
=
\lim_{\epsilon\to +0} 
\left[
\left( \int_0^{\frac{1}{y}-\epsilon} + \int_{\frac{1}{y}+\epsilon}^{u_m}\right)
 +
 \left( \int_{u_m}^{\frac{1}{x}-\epsilon} + \int_{\frac{1}{x}-\epsilon}^{1}\right)
 \right]
 .
\end{align}  
Of course, a calculation with, say, more than eight safe digits, will deserve an explicit control of the algorithmic details. 

Numerical examples for $F_1^v(d)$ are collected in Tables \ref{t-f1numer} and \eqref{t-f1numereps}.

\begin{table*}[tb]
\caption[]{\label{t-f1numer}
The Appell function $F_1$ of the massive vertex integrals as defined in \eqref{F1-002}.
As a proof of principle, only the constant term of the expansion in 
$d=4-2\varepsilon$ is shown, $F_1(1;1,\frac{1}{2};2;x,y)$.
Upper values: this calculation, \eqref{ss-f1vxy1}, lower values: \eqref{F1limit3}.
}
\renewcommand{\arraystretch}{1.2}
\begin{center}
\begin{tabular}{|l|l|ll|}
\hline
  $x -i\varepsilon_x$ & $y-i\varepsilon_y$ & $F_1(1;1,\frac{1}{2};2;x,y)$ &
\\
\hline\hline
$+11.1 - 10^{-12}i$ & $+12.1 - 10^{-12} i$ & 
   $-0.1750442480735 
   $&$      - 0.0542281294732 
   ~i$   
   \\ 
   & & 
   $-0.175044248073518778844982899126 $ & $ - 0.054228129473304027882097641167 ~i
$
\\ \hline  
$+11.1 - 10^{-12}i$   & $+12.1 + 10^{-12} i$  & $ +1.7108545293244 
$&$ + 0.0542281294732 
~ i $
   \\  
   & & $
+1.71085452932433557134838204175 $&$+ 0.05422812947148217381589270924 ~i
$
 \\  \hline  
 $+11.1 + 10^{-12}i$   & $+12.1 - 10^{-12} i$  & 
$ +1.7108545304114 
$&$ - 0.0542281294732 
~i$
  \\   
  & & $
 +
 1.71085452932433557134838204175 $&$- 0.05422812947148217381589270924 ~i$
 \\ \hline  
$+11.1 + 10^{-12}~i$   & $+12.1 + 10^{-12}~i$  
& 
  $-0.1750442480735 
  $&$ + 0.0542281294733 ~i$ 
 \\   
 & & $
 -0.175044248073518778844982899126  $&$+ 0.054228129473304027882097641167 ~i$
\\
\hline\hline
$+12.1 - 10^{-15} ~i$   & $+11.1 - 10^{-15} ~i$  &
 $-0.1700827166484 
 $&$ -0.0518684846037 
 ~i$ 
 \\ $+12.1 - 10^{-10} ~i$&  $+11.1 - 10^{-15} ~i$ & $
 -0.170082716648000581011657492792  $&$ - 0.05186848460465674976556525621 ~i $
 \\ \hline  
$+12.1 - 10^{-15} ~i$   & $+11.1 + 10^{-15} ~i$  &
 $-0.1700827166484 
 $ & $ - 1.7544202909955 
 ~i$
\\  
& & $
 -0.17008271664844025647268817399  $ & $- 1.75442029099557688735842562038 ~i$
 \\ \hline  
$+12.1 + 10^{-15} ~i$   & $+11.1 - 10^{-15} ~i$  &
 $ -0.1700827166484 
 $&$  + 1.7544202909955 
 ~ i$
  \\ 
  & & $
  -0.17008271664844025647268817399 $&$+ 1.75442029099557688735842562038 ~i$
 \\ \hline  
$+12.1 + 10^{-15}~ i$   & $+11.1 + 10^{-15} ~i$  &
 $-0.1700827166484 
 $&$ + 0.0518684846037 
 ~i$ 
 \\  $+12.1 - 10^{-10} ~i$ &  $+11.1 - 10^{-15} ~i$ & $
 -0.170082716648000581011657492792 $&$ + 0.05186848460465674976556525621 ~i$
\\
\hline\hline
  $+11.1 - 10^{-15}~i$   & $-12.1$   
 &  $-0.0533705146518 
 $&$ - 0.1957692111557 
 ~i$
 \\
  && 
 $-0.053370514651899444733494011521$ & $- 0.195769211155733985388920833693 ~i$
  \\ \hline  
  $+11.1 + 10^{-15}~i$   & $-12.1$   
 &  $-0.0533705146518 
 $&$ + 0.1957692111557 
 ~i$
 \\
  && 
 $-0.053370514651899444733494011521$ & $+ 0.195769211155733985388920833693 ~i$
\\
\hline\hline
  $-11.1                $   & $+12.1- 10^{-12}~i$  
&  $  +0.1060864084662 
$&$  - 0.1447440700082 
i$ 
\\  
& & $
+0.106086408476510642871335275994$ & $- 0.144744070021333407167349619088 ~i$
\\\hline
$ -11.1                $   &  $+12.1+ 10^{-12}~i$  
&  $   +0.1060864084662 
$&$  + 0.1447440700082 
i$ 
\\ 
& & $
+0.106086408476510642871335275994$ & $+ 0.144744070021333407167349619088 ~i$
\\\hline\hline
$ -12.1                $   &  $-11.1$  
&  $ +0.122456767687224028 
$  & 
\\
 && 
$+0.1224567676872240250651339516130 $&
\\\hline
\end{tabular}
\end{center}
\end{table*}

\begin{table*}[tb]
\caption[]{\label{t-f1numereps}
The Appell function $F_1(1-\epsilon;1,\frac{1}{2};2-\epsilon;x_c,y_c)$
as defined in \eqref{F1-002}, needed for $d=4-2\varepsilon$.
}
\renewcommand{\arraystretch}{1.2}
\begin{center}
\begin{tabular}{|l|l|ll|}
\hline
 $x -i\varepsilon_x$ & $y-i\varepsilon_y$ & $F_1(1-\epsilon;1,\frac{1}{2};2-\epsilon;x,y)$ 
 \\\hline
$ +11.1 - 10^{-12}i               $   &  $+12.1- 10^{-12}i$
&
$+(-0.17504424807358806571 $&$ -    0.05422812947328981004 i) $
\\&&
$+(- 0.00861885859131501092 $&$ -     0.39051763820462137566 i) \epsilon$
\\&&
$+ (+0.37518853545319785781 $&$ -     0.34047477405516524129 i ) \epsilon^2 $
\\&&
$+(+0.49765461883470790694 $&$ -     0.00717399489427550385 i ) \epsilon^3$ 
\\&&
$+( +0.32835724868237320395 $&$ +     0.23005850008124251183 i) \epsilon^4 $
\\&&
$+ (+0.11199125312340825478 $&$ + 
    0.25409725390712356585 i) \epsilon^5 $
    \\&&
 $  +( - 0.00954795237038318610$&$  + 
    0.17050760870656256341 i )\epsilon^6 $
    \\&&
 $  +( - 0.042178619945247575185$&$  + 
    0.085768627808384789724 i) \epsilon^7$
\\\hline
\end{tabular}
\end{center}
\end{table*}

\subsection{Numerical calculation of the box Appell function $F_1^b(d)$ \label{subs-f1vd} }
For the calculation of four-point Feynman integrals, one needs $F_1^b(d)$ as introduced in \eqref{F1-003}, both for $d=4$ and for $d=4+n-2\varepsilon$.
The box $F_1$-function is related to the vertex function $F_1^v(d)$ by \eqref{F1-003}.
Consequently, the numerics of the foregoing subsections may be taken over.

\section{The Lauricella-Saran function $F_s$ \label{app-subs-fs}}
For the calculation of the 4-point Feynman integrals, one needs the Lauricella-Saran function $F_S$ \cite{saran1955}.
Saran defines $F_S$ as three-fold sum \eqref{eq-fsdef}, see Eqn.~(2.9) in \cite{saran1955}.
He derives a {3-fold integral representation in Eqn.~(2.15) and a 2-fold integral in Eqn.~(2.16)}.
We will use the following quite useful representation, derived at p. 304 of \cite{saran1955}: 
\begin{align}\label{fsgeneral}
&
F_S(a_1,a_2,a_2;b_1,b_2,b_3;c,c,c,x,y,z)
\\
&=
\frac{\Gamma(c)}{\Gamma(a_1)\Gamma(c-a_1)}
\nonumber\\\nonumber
&
\int_0^1 dt \frac{t^{c-a_1-1} (1-t)^{a_1-1}}{(1-x+tx)^{b_1}}
F_1(a_2;b_2,b_3;c-a_1;ty,tz)
. 
 \end{align}
In our case, this becomes
\begin{align}\label{fsJ4}
& 
F_S^b(d) =
F_S\left(\frac{d-3}{2},1,1;1,1,\frac{1}{2};\frac{d}{2},\frac{d}{2},\frac{d}{2},x_c,y_c,z_c\right)
\nonumber\\
& 
=
\frac{\Gamma(\frac{d}{2})}{\Gamma(\frac{d-3}{2})\Gamma(\frac{3}{2})}
\\\nonumber
& \times
\int_0^1 dt \frac{\sqrt{t} (1-t)^{\frac{d-5}{2}}}{(1-x_c+x_c t)}
F_1(1;1,\frac12;\frac32;y_c t,z_c t)
\end{align}
Eqn. \eqref{fsJ4} is valid if $\Re e(d)>3$.
With a grain of care one may {often} use \eqref{fs1dim} for $F_1^S$.
Because the $F_1$ under the $t$-integral is finite and smooth, we have to concentrate only on the term $1/(1-x_c+x_c t)$, which develops a pole in the integration region at 
$t_x=(1-x)/x$ if $\Re e(x_c)=x>1$ and if $\Im m(x_c)=-\varepsilon_x$ is infinitesimal.

\subsection{Case (i) $F_S^{b}(d)$ at $x \leq 1$ \label{ss-f1yx1}}
For $x=1$, the integral \eqref{fsJ4} is not well-defined. 
If $x<1$, a direct, stable numerical integration of $F_S$ is trivial once $F_1$ is known.

\subsection{Case (ii) $F_S^{b}(d)$ at $x > 1$ \label{ss-f1yx2}}
If $x>1$, one has to apply a regularization procedure to $F_S^b(d)$, as it is described 
in \eqref{ss-f1vxy1}, and will get a stable result for $F_S$.
The calculation of the $F_1$ in the integrand in \eqref{fsJ4} is discussed in 
\ref{ss-dfour}.

One now has to study the singularity structure of the $t$-integral as a function of $x_c$ with regular $F_1^S$.
Introduce 
\begin{align}\label{eq-fsreg}
&
F_S^b(d) = \int_0^1 dt \frac{g_S(t) - g_S(t_x)}{1-x+x t}
+
 g_S(t_x) ~ I_S^{reg}(x_c) 
 ,
\end{align}
with 
\begin{align}
 g_S(t) = \sqrt{t} ~ (1-t)^{(d-5)/2} ~ F_1^{S}{(y_c t,z_c t)}
\end{align}
and 
\begin{align}
{ t_{x} = 1-\frac{1}{x}}
 .
\end{align}
The first integral in \eqref{eq-fsreg} is numerically stable, and what remains is to calculate analytically the integral 
\begin{align}
 I_S^{reg}(x_c) ={ +\frac{1}{x_c}~\int_0^1 \frac{dt}{t-t_{x_c}} 
 = 
 \frac{1}{x_c} \ln\left(1-\frac{1}{t_{x_c}}\right) 
 }
 .
\end{align}
For infinitesimal $\varepsilon_x$, we get
\begin{align}
 I_S^{reg}(x_c) \to
 \frac{1}{x} \left[ -\ln(x-1) +i\pi ~ \mathrm{sign}(\varepsilon_x) \right]
 . 
 \end{align}

\section*{Acknowledgements}
The authors would like to thank J. Bl{\"u}mlein for participation in an early stage of the project when first results on the vertex functions had been obtained.
T.R. would like to thank J.~Fleischer, J.~Gluza, M.~Kalmykov and O.Tarasov for helpful discussions.  
Khiem Hong Phan would like to thank J. Bl{\"u}mlein and DESY for the opportunity to work 
in 2015 and 2016 as a guest scientist at Zeuthen.
K.H.P's work is funded by Vietnam’s National Foundation for Science and 
Technology Development (NAFOSTED) under the grant number 103.01-2016.33.
The work of T.R. is funded by Deutsche Rentenversicherung Bund. 
T.R. is supported in part by a 2015 Alexander von Humboldt Honorary Research Scholarship of the Foundation for Polish Sciences (FNP) 
and by the Polish National Science Centre (NCN) under the Grant Agreement 
2017/25/B/\linebreak[4]ST2/01987.
Thanks are also due to J.~Usovitsch for assistance in numerical comparisons.

\bibliographystyle{elsarticle-num} 
\bibliography{2loops-1Loops} 

\end{document}